\documentclass[10pt, conference]{IEEEtran}
\usepackage{amsmath}
\usepackage{graphicx}
\usepackage{subfigure}
\usepackage{curves}
\usepackage{amsfonts}
\usepackage{amsmath,epsfig}
\usepackage{stfloats}
\usepackage{amssymb}
\usepackage{cite}
\usepackage{algorithm}
\usepackage{algorithmic}
\ifCLASSINFOpdf

\else

\fi

\begin{document}

\title{Polar Coding for Secure Transmission in MISO Fading Wiretap Channels}
\author{\IEEEauthorblockN{Mengfan Zheng, Meixia Tao, Wen Chen\\}\IEEEauthorblockA{Department of Electronic Engineering, Shanghai Jiao Tong University, China\\Email: \{zhengmengfan; mxtao; wenchen\}@sjtu.edu.cn\\}}
\maketitle

\begin{abstract}
  In this paper, we propose a polar coding scheme for secure communication over the multiple-input, single-output, single-antenna eavesdropper (MISOSE) fading channel. We consider the case of block fading channels with known eavesdropper channel state information (CSI) and the case of fading channels with known eavesdropper channel distribution information (CDI). We use the artificial noise assisted secure precoding method to maximize the secrecy capacity in the first case, and to overcome the unawareness of the eavesdropper CSI in the second case. We show that our proposed scheme can provide both reliable and secure communication over the MISOSE channel with low encoding and decoding complexity.
\end{abstract}

\section{Introduction}
  Providing physical-layer security in wireless communications has drawn much attention. Wyner proved in \cite{wyner1975wire} that both reliable and secure communication is possible through random coding, with the constraint that the eavesdropper channel is degraded with respect to the legitimate channel. Since then, a lot of work has been done on showing the existence of secure coding schemes for different kinds of channels. However, few of these results provide guidance for designing a specific polynomial-time coding scheme. Only for some special cases constructive solutions are proposed \cite{suresh2010strong}, \cite{thangaraj2007applications}, \cite{wei1991generalized}, \cite{cheraghchi2012invertible}.

  Polar codes \cite{arikan2009channel}, proposed by Arikan, have shown capacity-achieving property in both source coding and channel coding \cite{korada2009polar}. The structure of polar codes makes them suitable for designing wiretap codes. Polar codes for wiretap channels have been studied in \cite{andersson2010nested}, \cite{hof2010secrecy}, \cite{mahdavifar2011achieving} and \cite{koyluoglu2012polar}, and for relay-eavesdropper channels in \cite{duo2014secure}. It has been shown that polar codes achieve the secrecy capacity of the two-user, memoryless, symmetric and degraded wiretap channel. The secrecy capacity defines the maximum achievable rate of which there exists coding schemes that can guarantee both reliability and security in communications. To ensure a positive secrecy capacity, the eavesdropper channel must be degraded with respect to the legitimate channel. However, this condition may not be satisfied in practical environments. One intuition of solving this problem is to jam the eavesdropper channel while not affecting the legitimate channel. The multiple-antenna system makes this possible by injecting carefully designed artificial noise into the transmitting signal on purpose \cite{goel2008guaranteeing}, which is referred to as the artificial noise assisted secure precoding method.

  To the best of our knowledge, polar coding for secure transmissions in multiple-antenna systems has not been explored yet. In this paper, we consider the polar coding scheme for multiple-input, single-output, single-antenna eavesdropper (MISOSE) fading channels with binary inputs under two scenarios, the first is that the transmitter knows perfect and instantaneous channel state information (CSI) of both the desired receiver and the eavesdropper, which we refer to as the CSI case, and the second is that the transmitter knows perfect and instantaneous CSI of the legitimate channel, but only the channel distribution information (CDI) of the eavesdropper channel, which we refer to as the CDI case. We combine polar coding and the artificial noise assisted precoding together, using the artificial noise to maximize the secrecy capacity in the CSI case and to overcome the unawareness of the eavesdropper CSI in the CDI case. We analyse the secrecy capacity of the MISOSE channel with binary-input constraint after adding artificial noise, and find the optimal power allocation scheme between the information-bearing signal and the artificial noise to maximize the secrecy capacity. For the CSI case our polar coding scheme achieves the secrecy capacity, and for the CDI case we propose a scheme that guarantees a given secrecy rate is achievable with a given probability.

  The following notations will be used throughout this paper.
  Matrices and vectors are represented by bold upper case and bold lower case letters respectively. For a vector $\mathbf{b}$, $\mathbf{b}_i^j$ denotes the subvector $(b_i,...,b_j)$ of $\mathbf{b}$ and $\mathbf{b}_\mathcal{B}$ denotes the subvector $(b_i:i\in\mathcal{B})$ of $\mathbf{b}$, where $\mathcal{B}\subset\{1,...,N\}$ and $N$ is the length of $\mathbf{b}$. $\mathbf{F}^{\otimes n}$ denotes the $n^{th}$ Kronecker power of $\mathbf{F}$. $W(y|x)$ is the transition probability of a channel $W$. $Z \triangleq \sum_{y\in\mathcal{Y}}\sqrt{W(y|0)W(y|1)}$ is the Bhattacharyya parameter for the channel $W$. The operator $\oplus$ denotes the binary addition.

\section{Preliminaries on Polar coding}
  Polar codes \cite{arikan2009channel} are the first kind of codes that can achieve the capacity of a large class of channels with low encoding and decoding complexities. Polar codes are constructed based on a phenomenon called channel polarization.

    Let $\mathbf{G}=\mathbf{B}_N \mathbf{F}^{\otimes n}$, where $N=2^n$ is the code length, $\mathbf{B}_N$ is a permutation matrix known as bit-reversal and $\mathbf{F}=
      \begin{bmatrix}
        1 & 0 \\
        1 & 1
      \end{bmatrix}$
    is called the kernel matrix of a polar code. Consider $N$ independent uses of a binary- input discrete memoryless channel (B-DMC) $W$ as $N$ coordinate bit-channels. By applying the linear transformation $\mathbf{G}$ on these bit-channels we can get a new $N$-dimensional channel $W_N (y_1^N|u_1^N)$, and Arikan has proved that by observing the channels defined by the transition probabilities
    \begin{equation}
      W_N^{(i)}(\mathbf{y},\mathbf{u}_1^{i-1}|u_i)=\sum_{\mathbf{u}_{i+1}^N}{\frac{1}{2^{N-1}}
      W_N(\mathbf{y}|\mathbf{x})},
    \end{equation}
    these synthesized channels polarize in a way that, as $N$ goes large, $W_N^{(i)}$ tends to either a noiseless channel or a purely noisy channel. The idea of polar coding is to use those good polarized channels to transmit useful information (called information bits) and those bad ones to transmit some fix information (called frozen bits) known by both the transmitter and receiver in advance.

    Denote $\mathcal{A}$ as a subset of $\{1,...,N\}$ such that $\forall i\in\mathcal{A}$, $W_N^{(i)}$ is one of the noiseless polarized channels. Then the encoding procedure is given by
    \begin{equation}
    \label{eq:enc}
      \mathbf{x}_1^N=\mathbf{u}_{\mathcal{A}}\mathbf{G}_{\mathcal{A}} \oplus \mathbf{u}_{\mathcal{A}^C}
      \mathbf{G}_{\mathcal{A}^C}.
    \end{equation}

    If we set all frozen bits to be $0$, then (\ref{eq:enc}) reduces to
    \begin{equation}
    \label{eq:encs}
      \mathbf{x}_1^N=\mathbf{u}_{\mathcal{A}}\mathbf{G}_{\mathcal{A}}.
    \end{equation}

    The decoding algorithm of polar codes proposed by Arikan is successive cancellation (SC) decoding, defined by
    \begin{eqnarray}
    \label{eq:dec}
      \hat{u}_i=
        \begin{cases}
        u_i,  &  i\in\mathcal{A}^C \\
        h_i(y_i^N,\hat{u}_i^{i-1}),  &  i\in\mathcal{A}
        \end{cases},
    \end{eqnarray}
    where $h_i$ is the decision function defined as
    \begin{eqnarray}
    \label{eq:decf}
      h_i(y_i^N,\hat{u}_i^{i-1})=
        \begin{cases}
        0  &  \frac{W_N^{(i)}(\mathbf{y},\mathbf{u}_1^{i-1}|u_i=0)}{W_N^{(i)}(\mathbf{y},
        \mathbf{u}_1^{i-1}|u_i=1)}\geq 1 \\
        1  &  otherwise
        \end{cases}.
    \end{eqnarray}

    Arikan has proved in \cite{arikan2009channel} that the block error probability of a polar code under the SC decoding is upper bounded by $\sum_{i\in\mathcal{A}}{Z({W_N^{(i)}})}$, where $Z({W_N^{(i)}})$ is the Bhattacharyya parameter for $W_N^{(i)}$. In \cite{arikan2009rate} it is shown that for any $0<\beta<1/2$,
    \begin{equation}
    \label{eq:Z}
      \lim_{N\rightarrow\infty}{\frac{1}{N}|\{i\in[N]:Z({W_N^{(i)}})<\frac{1}{N}2^{-N^\beta}\}|}=I(W),
    \end{equation}
    where $I(W)$ is the symmetric capacity of $W$. For a given $\beta \in (0,1/2)$ and $\epsilon>0$, we can define the set of information indices as $\mathcal{A}=\{i\in[N]:Z({W_N^{(i)}})\leq\frac{1}{N}2^{-N^\beta}\}$. Then for sufficiently large $N$, the rate of
    \begin{equation}
    \label{ieq:r}
      R=\frac{\mathcal{A}}{N}\geq I(W)-\epsilon
    \end{equation}
    is achievable with block error probability upper bounded by
    \begin{equation}
    \label{ieq:ep}
      P_e \leq 2^{-N^\beta}.
    \end{equation}

    The key property of polar codes that makes them suitable for secure communication can be concluded into the following lemma \cite{korada2009polar}, which will be used in our code design:
    \newtheorem{polar}{Lemma}[section]
    \begin{polar}
    \label{lemma:Polar}
      Let $W_1$ and $W_2$ be two B-DMCs such that $W_2$ is degraded with respect to $W_1$. Denote the set of information indices of $W_i (i=1,2)$ after channel polarization by $\mathcal{A}_i$. Then $\mathcal{A}_1\supseteq \mathcal{A}_2$.
    \end{polar}

\begin{figure}[tb]
  \centering
  \includegraphics[width=6cm]{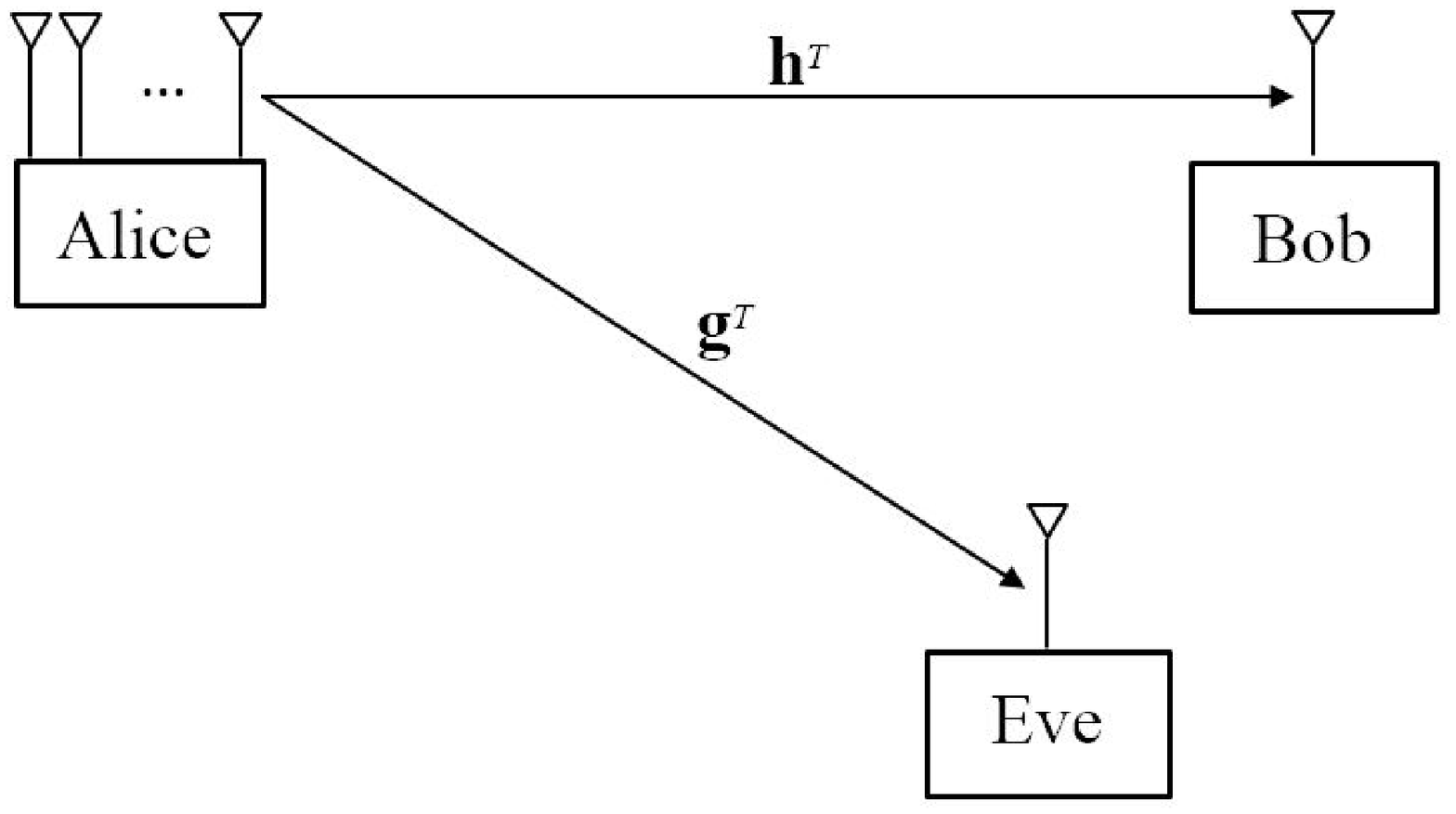}
  \caption{The MISOSE channel model.} \label{fig:SM}
\end{figure}
\section{System Model}
  \subsection{Channel Model}
  Fig. \ref{fig:SM} shows our system model. We consider the secure transmission in a MISOSE fading wiretap channel with a binary input alphabet. In this system, a transmitter (Alice), who is equipped with $N_A$ antennas ($N_A>1$), tries to communicate with a receiver (Bob) without leaking any confidential information to an eavesdropper (Eve). Bob and Eve each has a single antenna. The channel vectors of the legitimate channel and the eavesdropper channel are denoted by $\mathbf{h}^T$ and $\mathbf{g}^T$ respectively.

  The goal of designing secure coding schemes is to make communication both reliable and secure. To measure the degree of reliability and security we have the following metrics \cite{wyner1975wire}:

  The reliability condition:
  \begin{equation}
  \label{RC}
    \lim\limits_{N \to \infty}{\mathrm{Pr}\{\hat{\mathbf{U}} \neq \mathbf{U}\}}=0,
  \end{equation}

  And the security condition:
  \begin{equation}
  \label{SC}
    \lim\limits_{N \to \infty}{I(\mathbf{U};\mathbf{Z})/N}=0,
  \end{equation}
  where $N$ is the code length, $\mathbf{U}$ is the uncoded message, $\hat{\mathbf{U}}$ is Bob's decoding of $\mathbf{U}$, $\mathbf{Z}$ is Eve's observation of $\mathbf{X}$.

  \subsection{Artificial Noise Assisted Signal Model}
    As we know, a positive secrecy capacity can only be obtained if the legitimate channel has a better condition than the eavesdropper channel. In our system model, this means that the legitimate channel's signal-to-noise ratio (SNR) must be higher than the eavesdropper channel's. This condition can not always be guaranteed in practical scenarios. To solve this issue, the artificial noise (AN) scheme was proposed \cite{goel2008guaranteeing}. It is a jamming technique aimed at degrading the eavesdropper channel while not affecting the legitimate channel.

    The artificial noise assisted secure precoding method \cite{goel2008guaranteeing}, proposed by Goel and Negi, is a promising technique for secure communication over MIMO wiretap channels. The idea is to inject artificial noise into the null space of the legitimate channel so that only the eavesdroppers will be interfered. Let $\mathbf{h}^T=U\mathbf{\lambda V}^H$ be the singular value decomposition (SVD) of $\mathbf{h}^T$ and let $\mathbf{V}=[\mathbf{p},\mathbf{Z}]$, where $\mathbf{p}$ is the first column of $\mathbf{V}$. Then $\mathbf{p}$ is the orthonormal basis of $\mathbf{h}$, and $\mathbf{Z}$ is an orthonormal basis of $\mathbf{h}^T$'s null space. In the AN-aided scheme, the transmitted signal $\mathbf{x}$ is the sum of information-bearing signal $\mathbf{p}u$ and artificial noise signal $\mathbf{Zv}$,
    \begin{equation}
      \mathbf{x}=\mathbf{p}u+\mathbf{Zv}=\mathbf{V}
      \begin{bmatrix}
      u\\
      \mathbf{v}
      \end{bmatrix},
    \end{equation}
    where $u \in \{-\sqrt{\mathrm{P}_u},+\sqrt{\mathrm{P}_u}\}$ is the secret information bit, $\mathrm{P}_u$ is the power per information bit, and $\mathbf{v}$ is an artificial noise vector whose elements $v_i \ (i\in\{1,2,\cdots,N_A-1\})$ are i.i.d. Gaussian r.v.'s $v_i \sim \mathcal{N}_\mathbb{R}(0,\sigma_v^2)$. Let $\mathrm{P}_v=\sigma_v^2$ be the average power of an element in the artificial noise vector $\mathbf{v}$. Suppose the maximum transmitting power is $\mathrm{P}_t$, then $\mathrm{P}_u$ and $\mathrm{P}_v$ satisfy the power constraint of
    \begin{equation}
    \label{ieq:PC}
      \mathrm{P}_u+(N_A-1)\mathrm{P}_v\leq \mathrm{P}_t.
    \end{equation}

    Since $\mathbf{Z}\in \mathbb{R}^{N_A \times (N_A-1)}$ is an orthonormal basis of $\mathbf{h}^T$'s null space, we have $\mathbf{h}^T\mathbf{Z}=\mathbf{0}$. The signals received by Bob and Eve are then given by
    \begin{align}
      \label{eq:y}
      y&=\mathbf{h}^T\mathbf{x}+n_B=\mathbf{h}^T\mathbf{p}u+n_B,\\
      \label{eq:z}
      z&=\mathbf{g}^T\mathbf{x}+n_E=\mathbf{h}^T\mathbf{p}u+\mathbf{g}^T\mathbf{Zv}+n_E.
    \end{align}
    where $n_B$ and $n_E$ are additive white Gaussian noises (AWGN) of the legitimate channel and the eavesdropper channel respectively, $n_B \sim \mathcal{N}_\mathbb{R}(0,\sigma_B^2)$ and $n_E \sim \mathcal{N}_\mathbb{R}(0,\sigma_E^2)$. 
    Since $\|\mathbf{g}^T\mathbf{Z}\|$ is not necessarily 0 (unless $\mathbf{Z}$ happens to be an orthogonal basis of $\mathbf{g}^T$'s null space), Eve's channel is degraded by the artificial noise with high probability while Bob's channel remains unaffected.

    The SNR of the legitimate channel is given by:
    \begin{equation}
    \label{eq:SNRb}
      \mathrm{SNR_B}=\frac{\mathrm{P}_u\|\mathbf{h}^T\mathbf{p}\|^2}{\sigma_B^2}.
    \end{equation}

    The SNR of the eavesdropper channel:
    \begin{equation}
      \mathrm{SNR_E}=\frac{\mathrm{P}_u\|\mathbf{g}^T\mathbf{p}\|^2}{\mathbf{g}^T\mathbf{Z}
      \mathrm{E}[\mathbf{vv}^T]\mathbf{Z}^T\mathbf{g}+2\mathrm{E}[\mathbf{g}^T\mathbf{Zv}n_E]+\sigma_E^2}.
    \end{equation}

    Since the entries of the artificial noise vector $\mathbf{v}$ are i.i.d. Gaussian r.v.s, we have $\mathrm{E}[\mathbf{vv}^T]=\sigma_v^2 \mathbf{I}_{(N-1)\times(N-1)}$ and $\mathrm{E}[\mathbf{g}^T\mathbf{Zv}n_E]=0$. Thus
    \begin{equation}
    \label{eq:SNRe}
      \mathrm{SNR_E}=\frac{\mathrm{P}_u\|\mathbf{g}^T\mathbf{p}\|^2}{\mathrm{P}_v\|\mathbf{g}^T\mathbf{Z}\|^2+\sigma_E^2}.
    \end{equation}


    It is clear from (\ref{eq:SNRb}) and (\ref{eq:SNRe}) that by adjusting the information-bearing signal power $\mathrm{P}_u$ and the artificial noise power $\mathrm{P}_v$, we can make the eavesdropper channel worse than the legitimate channel, given that $\mathbf{g}$ is known to the transmitter and $\|\mathbf{g}^T\mathbf{Z}\|^2>0$.

\section{Secrecy Capacity Analysis}


  The purpose of adding AN into the channel is to jam the eavesdropper so that the capacity of the eavesdropper channel decreases and the secrecy capacity increases. The definition of instantaneous secrecy capacity is \cite{oggier2011secrecy}:
  \begin{equation}
    C_S \triangleq \max\limits_{p(u)}\{I(u;y)-I(u;z)\}.
  \end{equation}

  In our model the input $u$ is restricted to a binary alphabet $\mathcal{X}$ with uniform distribution, i.e., $p(u=-\sqrt{\mathrm{P}_u})=p(u=+\sqrt{\mathrm{P}_u})=1/2$. Thus the instantaneous secrecy capacity can be rewritten as
  \begin{equation}
  \label{eq:Cs}
    C_S \triangleq I(u;y)-I(u;z)=C_B-C_E,
  \end{equation}
  where $C_B$ and $C_E$ are the binary-input constraint channel capacities of the legitimate channel and the eavesdropper channel respectively. The mutual information $I(x;y)$ is defined as
  \begin{equation}
  \label{eq:I}
    I(x;y)=\sum\limits_{x}{\int\nolimits_{-\infty}^{+\infty}{p(y|x)p(x)\log_2{\frac{p(y|x)}{p(y)}}\mathrm{d}y}}.
  \end{equation}

  We assume the CSI of the legitimate channel is perfectly available at all nodes. Then the instantaneous capacity of the legitimate channel is
  \begin{equation}
  \label{eq:CB}
  \begin{split}
    C_B&=\frac{1}{2}\sum\limits_{u}{\int\nolimits_{-\infty}^{+\infty}{\frac{1}{\sqrt{2\pi}\sigma_B}}\exp{\Big{(}-\frac{(y-\mathbf{h}^T\mathbf{p}u)^2}{2\sigma_B^2}\Big{)}}}\\
       &\times \log_2{\frac{\exp{(-\frac{(y-\mathbf{h}^T\mathbf{p}u)^2}{2\sigma_B^2})}}{\frac{1}{2}\exp{(-\frac{(y-\mathbf{h}^T\mathbf{p}u)^2}{2\sigma_B^2})}+\frac{1}{2}
    \exp{(-\frac{(y+\mathbf{h}^T\mathbf{p}u)^2}{2\sigma_B^2})}}}\mathrm{d}y\\
    &=1-\frac{1}{\sqrt{2\pi}\sigma_B}\int\nolimits_{-\infty}^{+\infty}{\exp{\Big{(}-\frac{(y-\mathbf{h}^T\mathbf{p}\sqrt{\mathrm{P}_u})^2}{2\sigma_B^2}\Big{)}}}\\
    &~~~~~~~~~~~~~~~~\times\log_2{\Big{(}1+\exp{(-\frac{2\mathbf{h}^T\mathbf{p}\sqrt{\mathrm{P}_u} y}{\sigma_B^2})}\Big{)}}\mathrm{d}y.
  \end{split}
  \end{equation}

  In \cite{komo2006upper} the authors give a series expression for evaluating (\ref{eq:CB}), which is given by
  \begin{equation}
  \label{eq:CBs}
  \begin{split}
    C_B&=1-\frac{1}{\ln{2}}\Big{[}\frac{2\beta_B e^{-\beta_B^2 /2}}{\sqrt{2\pi}}-(2\beta_B^2-1)Q(\beta_B)\\
    &~~~~~~~~+\sum_{i=1}^\infty\frac{(-1)^{i-1}}{i(i+1)}e^{2i(i+1)\beta_B^2}Q[(2i+1)\beta_B]\Big{]},
  \end{split}
  \end{equation}
  where $\beta_B=\mathbf{h}^T\mathbf{p}\sqrt{\mathrm{P}_u}/\sigma_B=\sqrt{\mathrm{SNR_B}}$ and $Q(x)=\int_x^\infty \frac{1}{\sqrt{2\pi}}e^{-t^2/2}\mathrm{d}z$. When evaluating the channel capacity using (\ref{eq:CBs}), the maximum of $i$ does not need to be very large since $e^{2i(i+1)\beta_B^2}Q[(2i+1)\beta_B]$ vanishes fast with the increasing of $i$. For example, when $\beta_B=3$, the results when $\max\{i\}=3,4,5$ does not make much difference.

  The instantaneous channel capacity of the eavesdropper channel without adding artificial noise can be given in the same expression as (\ref{eq:CBs}), with $\beta_B$ replaced by $\beta_E=\mathbf{g}^T\mathbf{p}\sqrt{\mathrm{P}_u}/\sigma_E$. After adding artificial noise, since the elements of the artificial noise vector $\mathbf{v}$ are independent Gaussian r.v.'s, the total noise term in (\ref{eq:z}) $\mathbf{g}^T\mathbf{Zv}+n_E$ is also a Gaussian r.v. with zero mean and variance $\sigma_{\bar{E}}^2=\|\mathbf{g}^T\mathbf{Z}\|^2 \mathrm{P}_v+\sigma_E^2$. If the eavesdropper channel's CSI is known by the transmitter, the channel capacity can be similarly given by


    \begin{equation}
    \label{eq:CE}
    \begin{split}
      C_E&=1-\int\nolimits_{-\infty}^{+\infty}{\frac{1}{\sqrt{2\pi}\sigma_{\bar{E}}}}\exp{\Big{(}-\frac{(z-\mathbf{g}^T\mathbf{p}\sqrt{\mathrm{P}_u})^2}{2\sigma_{\bar{E}}^2}\Big{)}}\\
      &~~~~~~~~~~~~~~~~~~\times\log_2{\Big{(}1+\exp{(-\frac{2\mathbf{g}^T\mathbf{p}\sqrt{\mathrm{P}_u}z}{\sigma_{\bar{E}}^2})}\Big{)}\mathrm{d}z}\\
         &=1-\frac{1}{\ln{2}}\Big{[}\frac{2\beta_{\bar{E}} e^{-\beta_{\bar{E}}^2 /2}}{\sqrt{2\pi}}-(2\beta_{\bar{E}}^2-1)Q(\beta_{\bar{E}})\\
    &~~~~~~~~+\sum_{i=1}^\infty\frac{(-1)^{i-1}}{i(i+1)}e^{2i(i+1)\beta_{\bar{E}}^2}Q[(2i+1)\beta_{\bar{E}}]\Big{]},
    \end{split}
    \end{equation}
    where $\beta_{\bar{E}}=\mathbf{g}^T\mathbf{p}\sqrt{\mathrm{P}_u}/\sigma_{\bar{E}}=\sqrt{\mathrm{SNR_E}}$.

    The instantaneous secrecy capacity can then be obtained by applying (\ref{eq:CBs}) and (\ref{eq:CE}) to (\ref{eq:Cs}):
    \begin{equation}
    \label{eq:CsCSI}
    \begin{split}
      C_S &= C_B-C_E\\
          &= \mathrm{F}(\beta_{\bar{E}})-\mathrm{F}(\beta_B),
    \end{split}
    \end{equation}
    where
    \begin{equation}
    \begin{split}
    \mathrm{F}(x)&=\frac{1}{\ln{2}}\Big{[}\frac{2x e^{-x^2 /2}}{\sqrt{2\pi}}-(2x^2-1)Q(x)\\
                 &~~~~+\sum_{i=1}^\infty\frac{(-1)^{i-1}}{i(i+1)}e^{2i(i+1)x^2}Q[(2i+1)x]\Big{]}.
    \end{split}
    \end{equation}

    To maximize the secrecy capacity, we need to find the optimal power allocation scheme between the information-bearing signal and the artificial noise. It is obvious that the equation in (\ref{ieq:PC}) holds when the maximum secrecy capacity is obtained, otherwise just by simply allocating the redundant power to the artificial noise will make the secrecy capacity larger. The optimization problem can be formulated as
    \begin{align}
    \label{eq:PA}
    \max \quad &C_s=\mathrm{F}(\beta_{\bar{E}})-\mathrm{F}(\beta_B)\\
    \label{eq:PC}
    s.t.\quad &\mathrm{P}_u+(N_A-1)\mathrm{P}_v=\mathrm{P}_t\\
    &\mathrm{P}_u > 0  \\
    &\mathrm{P}_v \geq 0
    \end{align}

    This optimization problem can be turned into a single-variable problem due to the power constraint of (\ref{eq:PC}), and can be easily solved.
    \begin{figure}[tb]
    \centering
    \includegraphics[width=9.4cm]{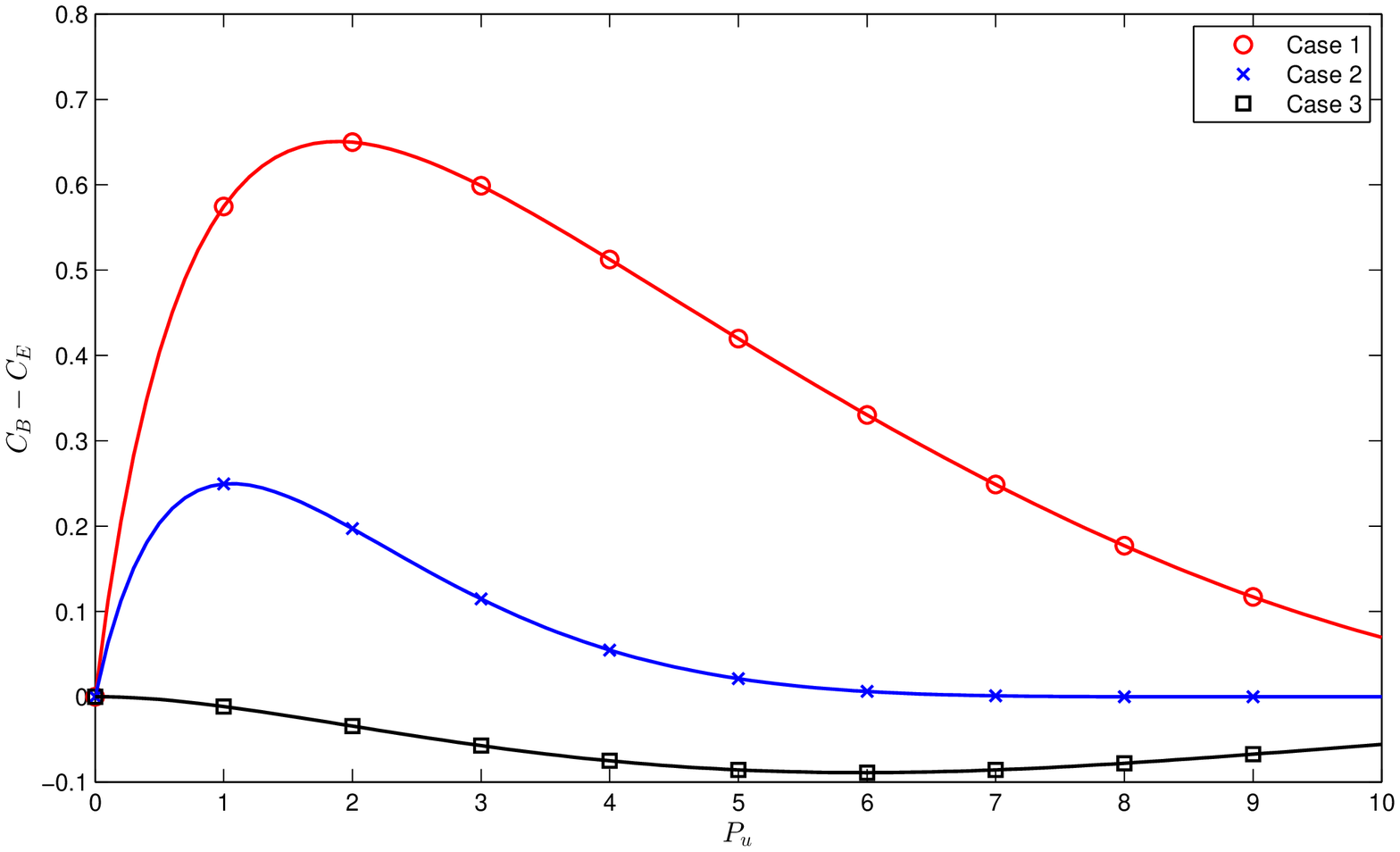}
    \caption{Secrecy capacity vs. $\mathrm{P}_u$ with $\mathrm{P}_t=10$.} \label{fig:CS}
    \end{figure}
    \begin{figure}[tb]
    \centering
    \includegraphics[width=9.4cm]{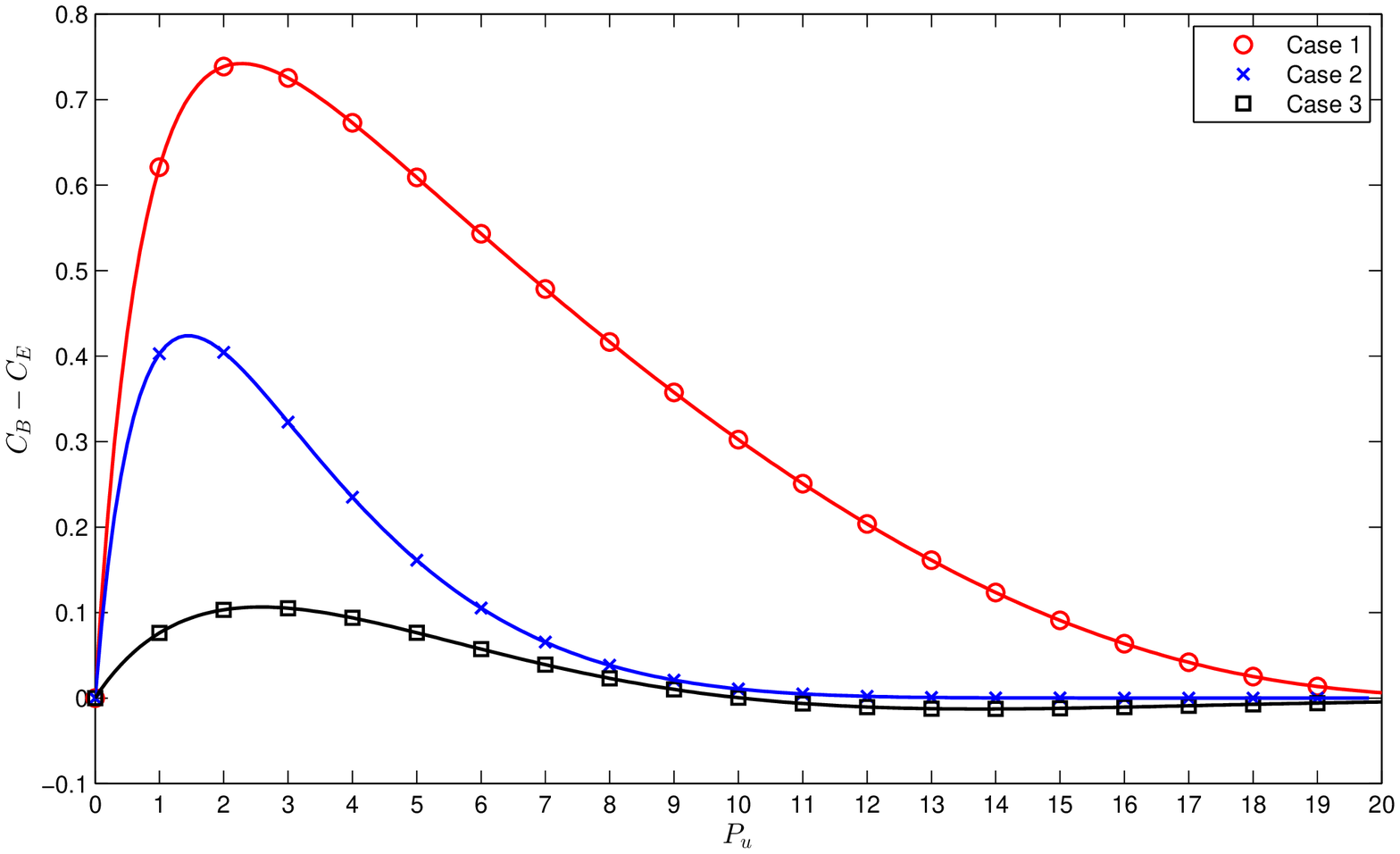}
    \caption{Secrecy capacity vs. $\mathrm{P}_u$ with $\mathrm{P}_t=20$.} \label{fig:CS1}
    \end{figure}

    Fig. \ref{fig:CS} shows three examples of the secrecy capacity change with different $\mathrm{P}_u$. The total transmitting power is $\mathrm{P}_t=10$. The channel vector pairs ($\mathbf{h}$, $\mathbf{g}$) in three cases are randomly generated, and $\sigma_B^2$ and $\sigma_E^2$ in each case are different.  




    It can be seen that for case 1 and 2, there exists an optimal $\mathrm{P}_u$ that maximizes the secrecy capacity. But for case 3, $C_B-C_E$ is always below 0, which indicates a 0 secrecy capacity. But if we increase the total transmitting power to 20, as is shown in fig. \ref{fig:CS1}, case 3 will also have an optimal solution with a positive maximum secrecy capacity.

\section{Polar Coding Scheme}
  As can be seen from (\ref{eq:y}) and (\ref{eq:z}), in the AN-assisted precoding scheme, the channel is equivalent to a single-input single-output wiretap fading channel. The equivalent channel gains of the legitimate channel and the eavesdropper channel are $h'=\mathbf{h}^T\mathbf{p}$ and $g'=\mathbf{g}^T\mathbf{p}$ respectively. Thus, the design of polar codes for MISOSE fading channels with AN-assisted precoding is similar to the polar coding design for fading wiretap channels. The coding procedure is done on the message bit $u$.


  Moreover, as we have assumed that perfect and instantaneous CSI of the legitimate channel is known at all ends, the legitimate channel is equivalent to a binary-input AWGN channel. We first discuss the case that the eavesdropper CSI is known by the transmitter (the CSI case). We show that in this case our polar coding scheme achieves the secrecy capacity with binary-input constraint. Then we discuss the case that the transmitter only has the CDI of the eavesdropper channel (the CDI case) and propose a feasible scheme that can guarantee a given secrecy rate is achievable with a given probability. This is a more practical assumption since the eavesdropper is passive and gives no feedback to the transmitter. However, due to the unawareness of the wiretap channel, perfect secrecy can not be achieved.

\subsection{Code Construction}
  \subsubsection{The CSI Case}
  In this case the legitimate channel and the eavesdropper channel are equivalent to two different binary-input AWGN channels. Thus the code design problem becomes the problem of designing polar codes for binary-input AWGN wiretap channels. As has been discussed in \cite{mahdavifar2011achieving}, the main idea of polar coding for the wiretap channel is to partition the set information indices of a polar code for the legitimate channel into two subsets: the secure bits and the insecure bits.

  Denote the set of information indices of polar codes for the legitimate channel by $\mathcal{A}$, and of the eavesdropper channel by $\mathcal{E}$. The corresponding sets of frozen indices are then $\mathcal{A}^C$ and $\mathcal{E}^C$ respectively. As the CSIs of both channels are known by the transmitter, the code construction can be determined using the Monte Carlo approach \cite{arikan2009channel} or a more accurate method \cite{tal2011construct}. Before the code design, we first optimize the power allocation as discussed in the previous section and determine the maximum secrecy capacity. In this way we can guarantee that the eavesdropper is degraded with respect to the legitimate channel even if the original eavesdropper channel is better. According to lemma \ref{lemma:Polar}, if the eavesdropper channel is degraded with respect to the legitimate channel, $\mathcal{E} \subseteq \mathcal{A}$ and $\mathcal{A}^C \subseteq \mathcal{E}^C$. Thus an $N$-bit long codeword can be partitioned into three subsets:
  \begin{align}
    \label{eq:G}
    \mathcal{G} &\triangleq \mathcal{A}-\mathcal{E}\\
    \label{eq:M}
    \mathcal{M} &\triangleq \mathcal{A} \cap \mathcal{E}\\
    \label{eq:B}
    \mathcal{B} &\triangleq \mathcal{A}^C
  \end{align}

\begin{figure}[tb]
    \centering
    \includegraphics[width=9cm]{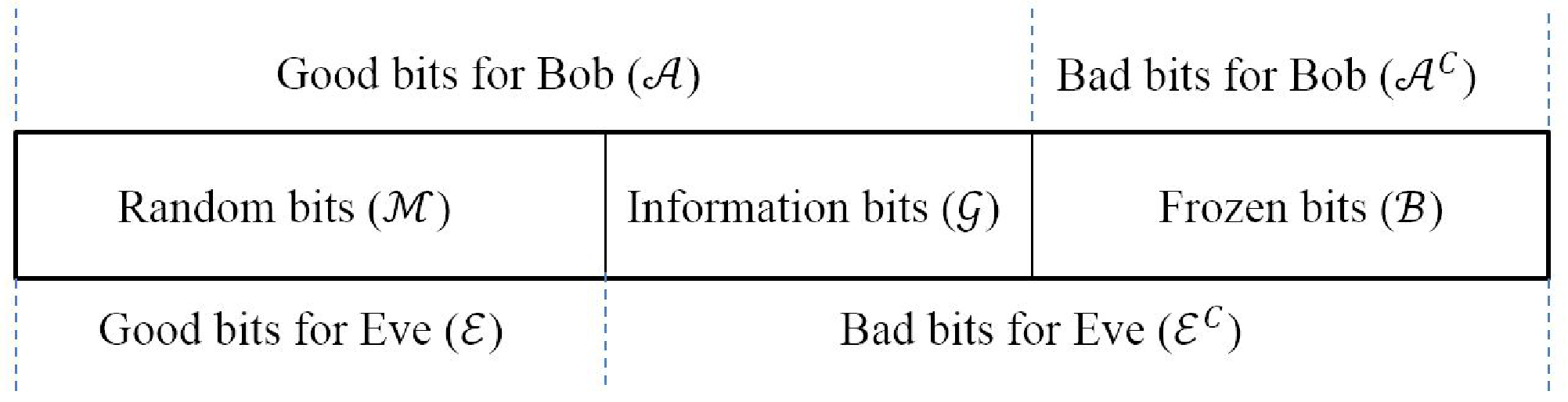}
    \caption{Order-permuted version of a codeword.} \label{fig:PC}
  \end{figure}

  Fig. \ref{fig:PC} gives a visualized view of this partition. It is obvious that $\mathcal{G}\bigcup\mathcal{M}\bigcup\mathcal{B}=\{1,...,N\}$. In a codeword, $\mathcal{G}$ corresponds to the bits that are good for Alice but bad for Eve, $\mathcal{M}$ corresponds to those good for both of them, and $\mathcal{B}$ corresponds to those bad for both of them. To ensure security, we fill those bits in $\mathcal{G}$ with confidential information, those in $\mathcal{M}$ with random bits to confuse the eavesdropper, and those in $\mathcal{B}$ with frozen bits. Let $|\mathcal{A}|=n$ and $|\mathcal{E}|=m$, then the code rate is defined as $R_S=|\mathcal{G}|/N=(n-m)/N$. From (\ref{eq:Z}) we have
  \begin{equation}
  \label{ieq:R}
  \begin{split}
    \lim_{N\rightarrow\infty}{R_S}&=\lim_{N\rightarrow\infty}{(|\mathcal{A}|/N-|\mathcal{E}|/N)}\\
    &=I(W_B)-I(W_E)\\
    &=C_S
    \end{split}
  \end{equation}
  This means the secrecy capacity can be achieved by our coding scheme.

  Due to the capacity-achieving property of polar codes, the reliability of our coding scheme is immediate. Since the information bits in our coding scheme are chosen from the information bits of the polar code for the legitimate channel, it is obvious from (\ref{ieq:ep}) that
  \begin{equation}
  \label{ieq:eps}
    \mathrm{Pr}\{\hat{\mathbf{U}} \neq \mathbf{U}\} \leq P_e \leq 2^{-N^\beta}
  \end{equation}
  where $P_e$ is the block error probability of polar codes with block length $N$. Thus our coding scheme satisfies the reliability condition in (\ref{RC}).

  It has been proved in \cite{mahdavifar2011achieving} that polar codes for the binary input symmetric degraded wiretap channel satisfies the security condition of (\ref{SC}). We will just skip the proof here.

  \subsubsection{The CDI Case}
  The CSI case is an ideal case that shows the theoretical limit of this coding scheme. If the transmitter does not have the instantaneous CSI of the eavesdropper channel, the security of communication will be more difficult to guarantee. In \cite{si2013polar} the author considered polar coding for fading channels without CSI at the transmitter, but with the constraint that the fading channel has only finite channel states, which is not suitable for our system.

  In this paper, we propose a feasible scheme to combat the unawareness of the eavesdropper CSI using artificial noise. To be more practical, we further assume that the variance of the eavesdropper channel noise is also unknown. In this case we drop the channel noise items in (\ref{eq:SNRe}), which corresponds to the worst case for secure communication that the eavesdropper channel is noiseless. From (\ref{eq:SNRe}) we have
  \begin{equation}
  \label{eq:SNRes}
    \mathrm{SNR_E}=\frac{\mathrm{P}_u\|\mathbf{g}^T\mathbf{p}\|^2}{\mathrm{P}_v\|\mathbf{g}^T\mathbf{Z}\|^2}.
  \end{equation}

  First, we evaluate the cumulative distribution function (CDF) $\Phi(\eta)$ of $\eta=\frac{\|\mathbf{g}^T\mathbf{p}\|^2}{\|\mathbf{g}^T\mathbf{Z}\|^2}$ numerically according to the distribution of $\mathbf{g}$, and find the value $\eta_0$ such that $\Phi(\eta_0)=p_0$ for a given probability $p_0$ close to 1. This indicates that
  \begin{equation}
  \label{eq:PrV}
    \mathrm{Pr}\Big{\{}\frac{\|\mathbf{g}^T\mathbf{p}\|^2}{\|\mathbf{g}^T\mathbf{Z}\|^2} \leq \eta_0\Big{\}}=p_0.
  \end{equation}

  Then, we do the power allocating optimization of (\ref{eq:PA}) with $\mathrm{SNR_E}=\mathrm{P}_u\eta_0/\mathrm{P}_v$, and find the optimal $\mathrm{P}_u=\mathrm{P}_{u,o}$ and the maximum secrecy capacity $C_{s,o}$. The equation of (\ref{eq:PrV}) and the noiseless assumption guarantee that a target secrecy capacity of $C_{s,o}$ can be achieved with probability at least $p_0$.

  Finally, we design polar codes in the same way as the CSI case.

\subsection{Encoding and Decoding}
  \textbf{Encoding Procedure}:

  The encoder first accepts an message sequence $\mathbf{s}$ of $n-m$ bits and a random bit sequence $\mathbf{r}$ of $m$ uniformly distributed binary bits, and generates the encoded sequence $\mathbf{u}$ by

  \begin{equation}
  \label{eq:Menc}
    \mathbf{u}_1^N=\mathbf{s}\mathbf{G}_{\mathcal{G}} \oplus \mathbf{r}\mathbf{G}_{\mathcal{M}}
  \end{equation}

  The frozen bits are set to all zeros.

  \textbf{Decoding Procedure}:

  The successive cancellation decoding algorithm requires to calculate the likelihood ratios (LR) of $L_N^{(i)}(\mathbf{y},\mathbf{u}_1^{i-1}|u_i)=\frac{W_N^{(i)}(\mathbf{y},\mathbf{u}_1^{i-1}|u_i=0)}{W_N^{(i)}(\mathbf{y},\mathbf{u}_1^{i-1}|u_i=1)}$, which are obtained recursively from $L_1^{(1)}(y_i)=p(y_i|0)/p(y_i|1)$ by Eq. (75) and (76) in \cite{arikan2009channel}. In the case of the equivalent binary-input AWGN channel of (\ref{eq:y}), $L_1^{(1)}(y_i)$ is given by
  \begin{equation}
  \label{eq:L}
  \begin{split}
    L_1^{(1)}(y_i)&=\frac{\frac{1}{\sqrt{2\pi}\sigma_B}e^{-\frac{(y_i+\mathbf{h}^T\mathbf{p}\sqrt{\mathrm{P}_u})^2}{2\sigma_B^2}}}{\frac{1}{\sqrt{2\pi}\sigma_B}e^{-\frac{(y_i-\mathbf{h}^T\mathbf{p}\sqrt{\mathrm{P}_u})^2}{2\sigma_B^2}}}\\
    &=e^{-\frac{2y_i \mathbf{h}^T\mathbf{p}\sqrt{\mathrm{P}_u}}{\sigma_B^2}}
  \end{split}
  \end{equation}

  The decoder first tries to recover both $\mathbf{s}$ and $\mathbf{r}$ together under the normal SC decoding procedure, then it extracts the secret message $\mathbf{s}$ according to the code construction.

  The complexities of polar encoding and decoding are both $O(n\log n)$ \cite{arikan2009channel}. This means our scheme also has low encoding/decoding complexities of $O(n\log n)$.

\section{Simulation Results}
  In this section we give some simulation results to show the performance of our coding scheme.

  \subsection{The CSI Case}

  For the CSI case, we assume that both the legitimate channel and the eavesdropper channel are with Rayleigh distribution of scale parameter 1. $\sigma_B^2$ is normalized to 1 and $\sigma_E^2$ is also set to 1. The maximum total transmitting power is $\mathrm{P}_t=3$. We randomly generate 100 channel vector pairs of ($\mathbf{h}$, $\mathbf{g}$), and for each channel vector pair, we simulate transmitting 1,000 information bits for each code length (from $2^4$ to $2^{13}$). Since there are 100 channel vector pairs to be simulated, the total number of transmitted information bits is 100,000 for each code length, which is sufficient to show a minimum bit error rates (BER) of $10^{-5}$. Given a channel vector pair, we first find the optimal information-bearing signal power and compute the corresponding legitimate channel capacity $C_B$ and eavesdropper channel capacity $C_E$. Then we calculate the polar code construction for the legitimate channel of rate $R_B=C_B-\delta$ and the polar code construction for the eavesdropper channel of rate $R_E=C_E$. We choose $\delta=0.11$ to ensure reliability for the legitimate channel since the code length we use is not very long. Finally we determine the secure polar code construction by (\ref{eq:G}), (\ref{eq:M}) and (\ref{eq:B}).

\begin{figure}[tb]
    \centering
    \includegraphics[width=9.4cm]{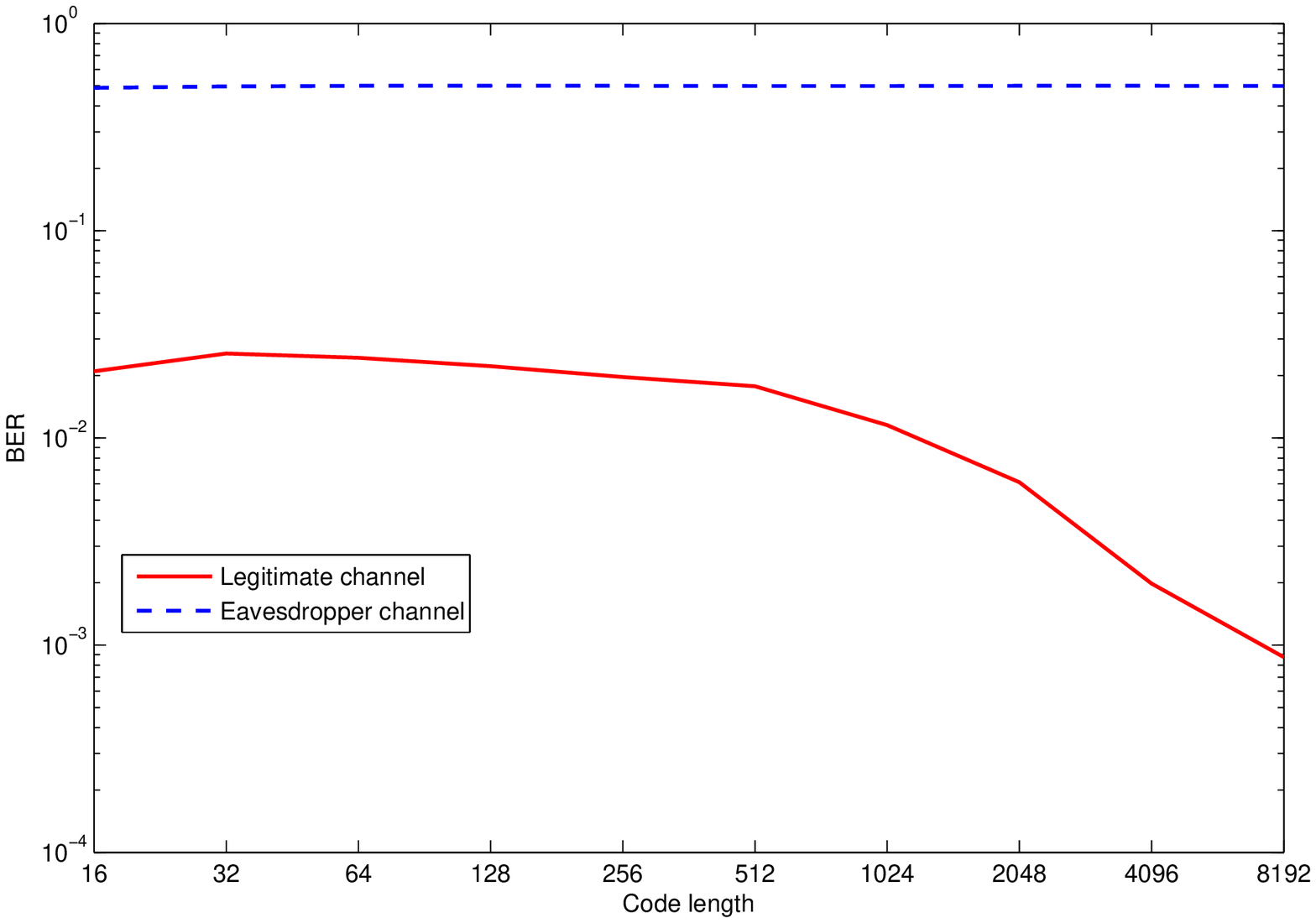}
    \caption{BER performance of the CSI case.} \label{fig:S1}
  \end{figure}

  After the simulation, we calculate the average BERs of the legitimate channel and the eavesdropper channel of different code lengths. Fig. \ref{fig:S1} shows the result. We can see that with the increasing of the code length, the BER of the legitimate channel decreases while the BER of the eavesdropper channel keeps almost invariant at about 0.5. The poor performance for short codes is due to the insufficiency of channel polarization since the polarized channels for information bits are comparatively less noiseless than channels for random bits.

  \begin{figure}[tb]
    \centering
    \includegraphics[width=9.4cm]{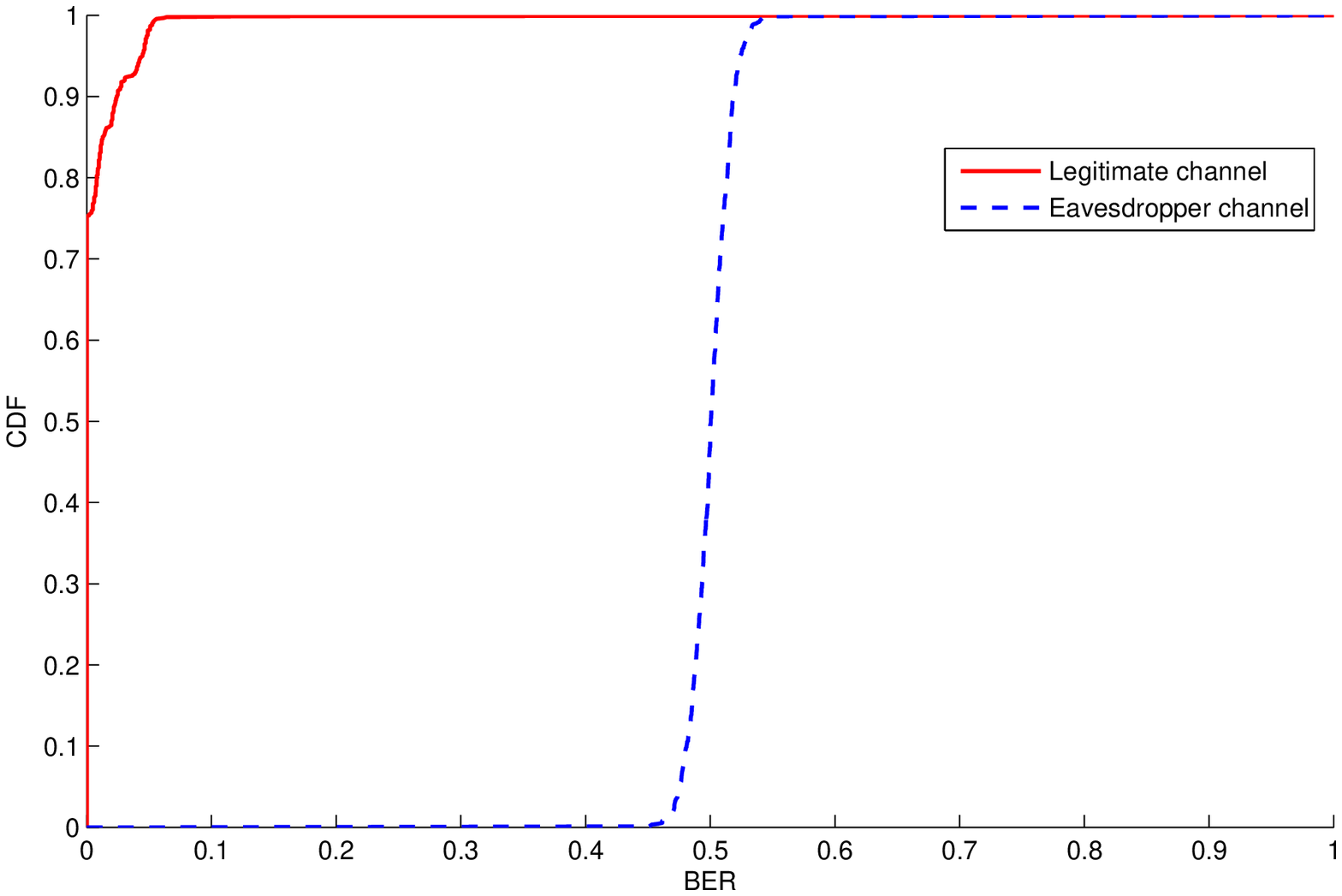}
    \caption{The distribution of BER for the CDI case.} \label{fig:CDIsim}
  \end{figure}

  \subsection{The CDI Case}

  For the CDI case, we assume the same distributions as the CSI case for both channels. The noise variances are $\sigma_B^2=\sigma_E^2=1$, and only $\sigma_B^2$ is known by the transmitter. The maximum total transmitting power is $\mathrm{P}_t=5$. We first randomly generate 30 $\mathbf{h}$s and design a secure polar code for each $\mathbf{h}$ with code length 4,096 and target probability $p_0=0.85$. The gap between the code rate for the legitimate channel and the channel capacity is set to $\delta=0.14$. For each $\mathbf{h}$ we randomly generate 30 $\mathbf{g}$s and evaluate the BER for each ($\mathbf{h}$, $\mathbf{g}$) pair by transmitting 1,000 information bits. After the simulation, we draw the CDF curves of the BERs.

  Fig. \ref{fig:CDIsim} shows the result. It can be seen that BER samples of the eavesdropper channel gather around 0.5, ranging from 0.45 to 0.55, and the vast majority of BER samples of the legitimate channel are 0. The average BER for the legitimate channel is $6.04\times 10^{-3}$ and for the eavesdropper channel $5.01\times 10^{-1}$. By increasing the target probability $p_0$ the CDF curve for the eavesdropper channel will be even steeper, and by increasing the code length or $\delta$ the reliability of the legitimate channel can be further enhanced.


\section{Conclusion}
In this paper we propose a secure coding scheme for the binary-input MISOSE fading channel based on polar codes. We first consider the case of block fading channel with perfect and instantaneous CSI and show that our scheme achieves the secrecy capacity. Then we further consider the case of fading channel with only CDI of the eavesdropper channel and propose a feasible method to ensure a given secrecy rate with a given probability. The artificial noise assisted secure precoding method is used in our scheme to degrade the eavesdropper channel. Our proposed scheme can provide both reliable and secure communication over the MISOSE channel with low encoding and decoding complexity.

\bibliographystyle{ieeetr}
\bibliography{AN-PolarCode}

\end{document}